\documentclass{ckm}                 
\usepackage{multirow}
\usepackage{txfonts}            

\confname{Workshop on the CKM Unitarity Triangle, IPPP Durham, April
  2003}

\title{CKM Reach at Hadronic Colliders}

\author{U Egede\thanks{On behalf of the \lhcb\ collaboration.}}

\address{Imperial College London, London SW7~2AZ, United Kingdom.}

\input ./symbols

\begin{document}

\begin{abstract}
  The analysis of the CKM parameters will take a leap forward when the
  hadronic $B$ factories receive their first data. I describe the
  challenges faced by $B$-physics at hadronic colliders and the
  expected reach in specific channels for the \lhcb, \btev, \atlas\ and
  \cms\ experiments.
\end{abstract}

\maketitle


\section{Introduction}
\label{sec:intro}
With the hadronic \B factories currently under construction or in the
design phase \B physics will enter a new era. The \lhcb, \btev,
\atlas\ and \cms\ experiments all have a time scale for \B physics
from 2007 onwards. The aim of the experiments is to gain a
comprehensive understanding of the CKM matrix for discovering physics
beyond the Standard Model. The much larger statistics and the access
to \Bs decays will allow to many cross checks of \CP violation that
are not possible at the current \B factories.

\begin{figure}[htbp]
  \centering
  \includegraphics[width=0.87\linewidth]{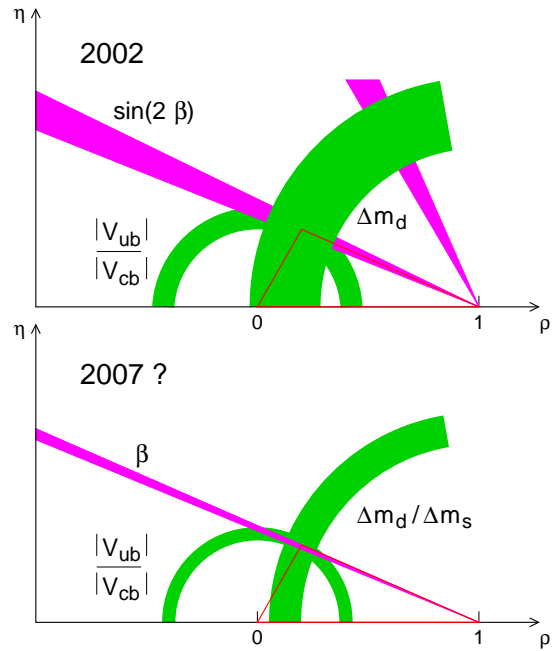}
  \caption{The anticipated improvement in three key measurements of
    the unitarity triangle between 2002 and 2007.}
  \label{fig:improvement}
\end{figure}

By 2007 the current \epem \B factories will have collected samples of
the order of $10^9$ \B-meson decays. This, combined with the data from
the Tevatron will give a precision on the value of the CKM angle
$\beta$ of $\sigma(\sin 2\beta) = \order(10^{-2})$ which is close to
the systematic uncertainty from penguin pollution in the channel $\B
\to \jpsi \KS$. At the same time the anticipated measurement of \Bs
mixing will improve the value of $\Vtd/\Vts$ from the partial
cancellation of the \Bd and \Bs form factors. The anticipated
improvement in the measurement of the apex of the unitarity triangle
between today~\cite{Lubicz:2003} and 2007 is shown in
figure~\ref{fig:improvement}.

In table~\ref{tab:conditions} the experimental conditions for the
different experiments are summarised and compared to a conceptual
design for a future \epem \B factory. Several points are worth further
comments:
\begin{itemize}
\item At a hadron collider the ratio between the \bbbar cross section
  and the total inelastic cross section is very small. The ratio
  improves as the centre of mass energy increases thus giving the LHC
  experiments an initial advantage compared to \btev\ running at a
  lower CM energy. It should also be noted that there are significant
  uncertainties in the \bbbar cross section at the LHC energy.
\item The large production of \Bs, $\Lambda_b$ and $B_c$ will open up
  entirely new areas of \B physics where the present data samples are
  very limited.
\item \lhcb\ will tune its luminosity to a level where most recorded
  events will have a single interaction. \btev's strategy is instead
  to cope with multiple events in a single bunch crossing by spreading
  them out across a larger vertex region where the individual primary
  vertices can be clearly identified.
\item The \atlas\ and \cms\ experiments do not have \B physics as
  their primary goal and will as such have a much lower trigger
  bandwidth dedicated to \B physics.
\end{itemize}
\begin{table*}[htbp]
  \centering
  \begin{tabular*}{\linewidth}[c]{@{\extracolsep{\fill}}lcccc}
                    & \multicolumn{2}{c}{\lhc}   
                            & \btev               & Super \babar \\
  \hline
  Beam type         & \multicolumn{2}{c}{\proton-\proton}
                            & \proton-\antiproton & \epem       \\
  Status            & \multicolumn{2}{c}{Construction}
                            & Pending finance     & Concept     \\
  $\sqrt{s}$        & \multicolumn{2}{c}{14~\tev} 
                            & 2~\tev              & 10.58~\gev  \\
  $\sigma_{\bbbar}$ & \multicolumn{2}{c}{500~\mub}
                            & 100~\mub            & 1.1~\nb     \\
  $\sigma_{\ccbar}$ & \multicolumn{2}{c}{3.5~\mbarn}
                            & 1~\mbarn            & 1.3~\nb     \\
  $\sigma_{\rm{inclusive}}$ 
                    & \multicolumn{2}{c}{80~\mbarn}
                            & 60~\mbarn            &      \\
  \Bp/\Bd/\Bs/$\Lambda_b$ mixture
                    & \multicolumn{2}{c}{40/40/12/8}
                            & 40/40/12/8           & 50/50/0/0  \\
  Bunch separation  & \multicolumn{2}{c}{25~\ns}
                            & 132/396~\ns          &            \\
  Size of collision region
                    & \multicolumn{2}{c}{5.3~\cm}
                            & 30~\cm               &            \\
  \hline
  \phantom{x} \\
                    & \lhcb              & \atlas/\cms
                            &    \btev             &  Super \babar \\
  \hline
  Pseudorapidity coverage      
                    & 2.1--5.3           & -2.5--2.5
                            & 2.1--5.3             &             \\
  \lum [\cmsqs]     & $2\times 10^{32}$  & $10^{33} (10^{34})$
                            & $2\times 10^{32}$    & $10^{36}$    \\
  $<n>$ per bunch crossing
                    & 0.5                & 2 (20)
                            & 1.6/4.8    &                        \\
  $n_{\bbbar}$ per $10^7$~s
                    & $10^{12}$          & $5\times 10^{12(13)}$
                            & $2\times 10^{11}$    & $10^{10}$   \\

  \end{tabular*}
  \caption{A comparison of the beam parameters and detector coverage
    for the hadronic \B factories and a comparison with a conceptual
    future \epem collider.}
  \label{tab:conditions}
\end{table*}

\section{Detector layout}
\label{section:layout}
For most studies of \CP violation in \B-meson decays we need to
identify the flavour of the \B-meson at production time. The dominant
contribution to this flavour tagging is through identification of
particles from the decay of the other \B-hadron created in the event.
Hence the detector needs to be designed such that a significant part
of the produced pairs of \B-hadrons both end up within the detector
acceptance. The most cost effective solution to this is to make a
detector that sits as much in the forward region as technology
allows. As both \B-hadrons tend to be boosted in the same direction
there is no synergetic effect from covering both forward regions. This
leads to the design of the \lhcb\ and \btev\ detectors as single-armed
forward spectrometers.

\begin{figure*}[htbp]
  \centering
  \includegraphics[height=\linewidth, angle=-90]{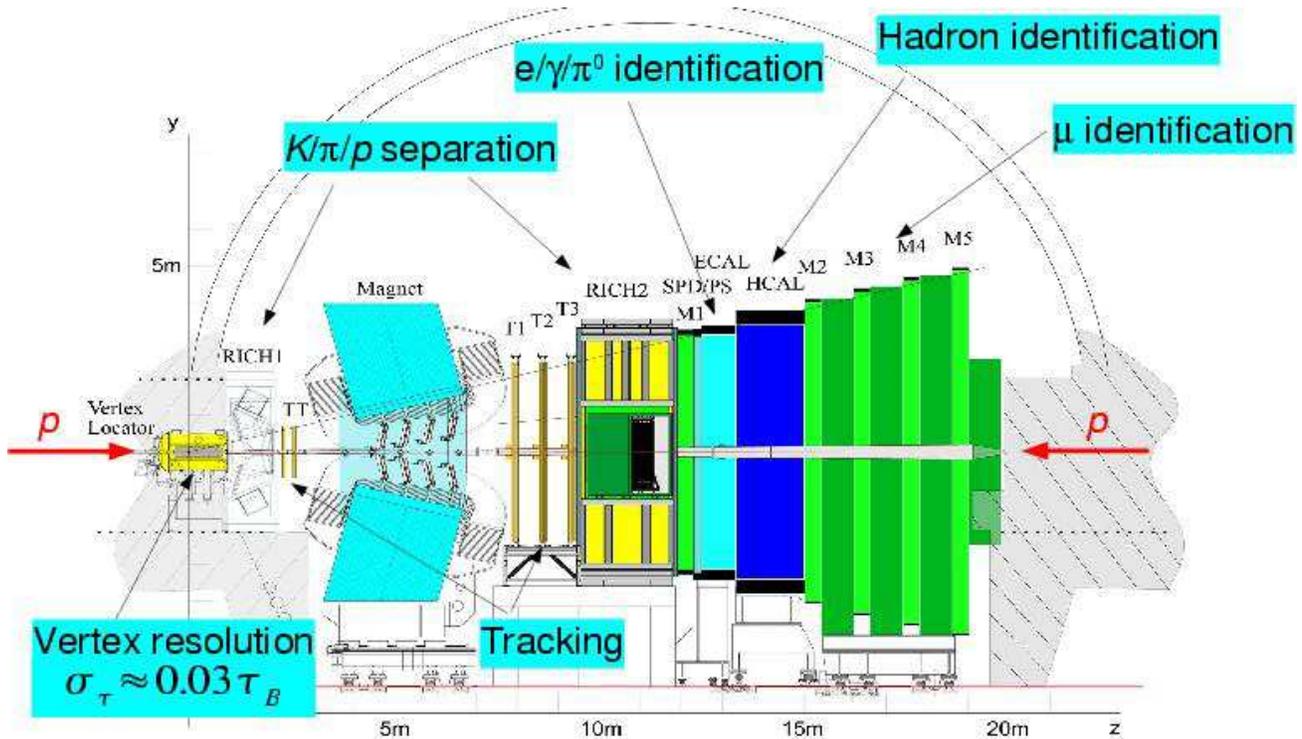}
  \caption{The design of the \lhcb\ detector. The collision point for
    the protons is within the vertex detector to the left in this
    drawing.}
  \label{fig:lhcb}
\end{figure*}

The overall design of the \lhcb\ detector is shown in
figure~\ref{fig:lhcb}. The most essential parts of the detector are: the
trigger system which reduces the rate of events going to mass storage
to an acceptable level; the vertex detector which provides the trigger
with secondary vertex identification and the physics with the ability
to resolve $\Bs$ oscillations; and the particle identification system
which provides the essential pion-kaon separation required for \CP
violation studies.

\subsection{Trigger}
\label{sec:trigger}
The single most demanding task for the hadronic \B physics experiments
will be the trigger. The combination of a cross section minimum bias
which is orders of magnitude larger than the \b cross section, with
the rare \B decays which are of interest requires a sophisticated
trigger that can suppress rates by many orders of magnitude. With a
rate of around $10^{12}$ \B-hadrons produced in a year the trigger
also have to be selective. This is a completely different situation to
current \epem colliders where all \B decays are recorded.

There are three main elements that allow identification of events with
a \B-hadron:
\begin{itemize}
\item Large transverse energy or momenta with respect to the beam
  axis. This is simply an indicator of a high mass particle decaying.
\item Vertices which are displaced from the primary vertex. This takes
  advantage of the long lifetime of \B-hadrons compared to other
  hadrons produced (\KS and $\Lambda$ live for much longer and do not
  interfere with the trigger).
\item High energy leptons either from semi-leptonic \B-decays or in
  pairs from \B-hadrons with a \jpsi in the decay chain. This will
  also be the trigger for rare $\B \to mupm$ decays.
\end{itemize}
The trigger strategy for \lhcb\ and \btev\ are in many ways quite
different.

In \lhcb\ the aim of the first trigger level is to identify events
with particles of high transverse energy or momentum with respect to
the beam axis. The high \pt and \et arises from the decay of high mass
objects and thus favours \B-hadrons to the background events with
lower mass hadrons. In addition these are the type of \B mesons events
that the final selection of events for physics analysis favour. At the
next trigger level secondary vertices are identified by placing cuts
on the significance of the impact parameter for tracks with respect to
the identified primary vertex. The last trigger identify more specific
classes of \B-decays using the results of the online reconstruction.

On the other hand the aim of the \btev\ trigger is to identify all
\B-hadron events and write them to tape. This is done by identifying
secondary vertices by placing cuts on the significance of the impact
parameter for tracks with respect to the identified primary vertex at
the full 7.6~\mhz rate of the beam crossings and then at the next
trigger level perform full secondary vertex reconstruction. The
procedure will intentionally also identify many charm events which
are predicted to take up 50\% of the rate of events going to tape.  We
show a summary and comparison of the \lhcb\ and \btev\ triggers in
table~\ref{tab:triggerA}.

The trigger for \atlas\ and \cms\ is much tighter as only a 10~\hz
output rate is allocated for \B physics in the first three years of
low-luminosity running for the two experiments. Both experiments rely
on a single muon of high \pt to trigger at the first level. The
identification of muon pairs from either $\B \to \jpsi X$ decays or
rare two-body decays to a pair of muons form the majority of the
trigger. A summary of the trigger is given in
table~\ref{tab:triggerB}.

More details can be found in the talks of
M.~Ferro-Luzzi~\cite{Ferro-Luzzi:2003}, L.~Moroni~\cite{Moroni:2003}
and A.~Starodumov~\cite{Starodumov:2003} at this workshop.

\begin{table}[htbp]
  \centering
  \begin{tabular*}{\linewidth}[c]{@{\extracolsep{\fill}}lcc}
     trigger type       & \lhcb      & \btev    \\
     \hline
     High \pt, high \et & 10~\mhz    &          \\ 
     Impact parameter   & 1~\mhz     & 7.6~\mhz \\
     Decay topology     &            & 80~kHz   \\
     Physics algorithms & 40~\khz    &          \\
     To mass storage    & 200~\hz    & 4~\khz   \\
  \end{tabular*}
  \caption{A simplified comparison of the trigger levels at \lhcb\
     and \btev. The 10~\mhz ingoing rate for \lhcb\ corresponds to the
     rate of bunch crossings with a visible interaction. In addition
     to what is given above both experiments have a dedicated first
     level trigger for events with one or two high \pt muons.}
  \label{tab:triggerA}
\end{table}
\begin{table}[htbp]
  \centering
  \begin{tabular*}{\linewidth}[c]{@{\extracolsep{\fill}}lcc}
     Trigger type       & \atlas              & \cms                  \\
     \hline
     Muon trigger       & 40~\mhz             & 40~\mhz               \\
     $\jpsi \to \ellell$, $\Ds \to \phi \pip$ & 20~\khz               \\
     Physics algorithms & 1~\khz              & 4~\khz                \\
     To mass storage    & 10~\hz              & 10~\hz                \\
  \end{tabular*}
  \caption{A simplified comparison of the \B physics trigger levels at \atlas\
     and \cms.}
  \label{tab:triggerB}
\end{table}

\subsection{Particle identification}
\label{sec:pid}
\begin{figure}[htbp]
  \centering
  \includegraphics[height=\linewidth, angle=-90]{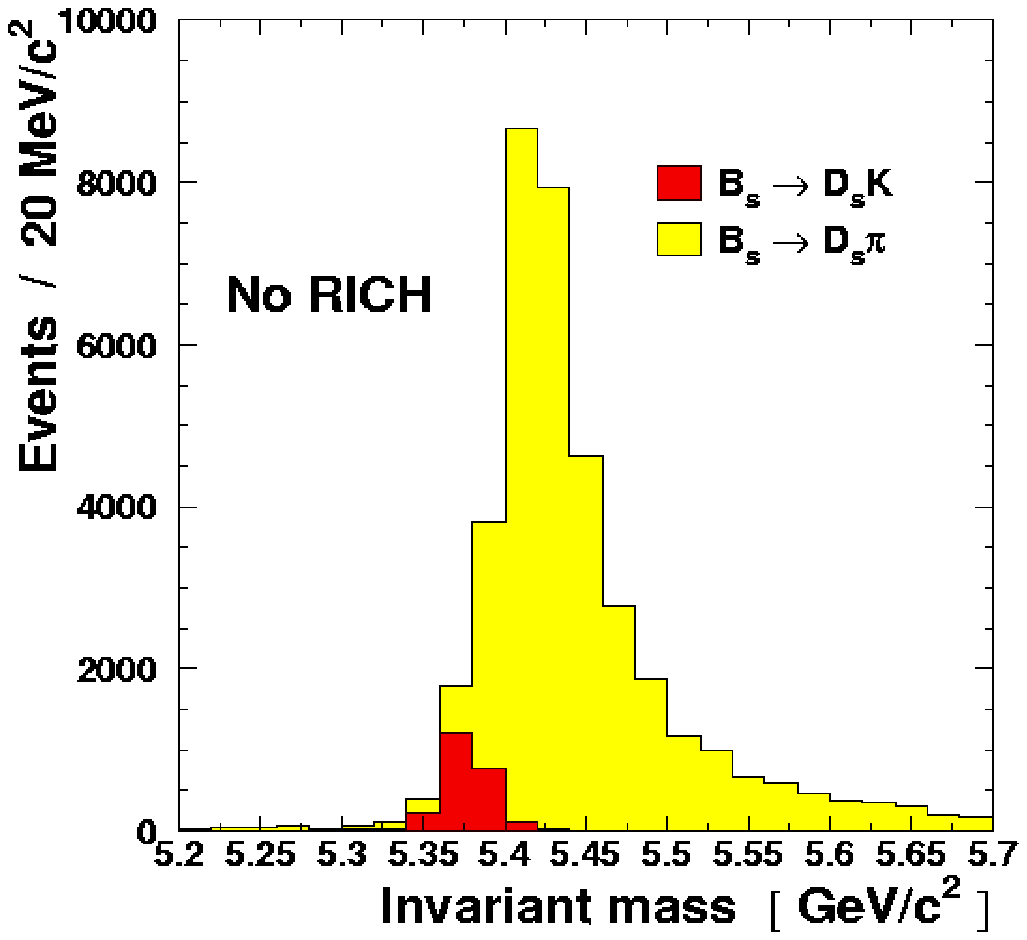}
  \vspace*{0.5cm}

  \includegraphics[height=\linewidth, angle=-90]{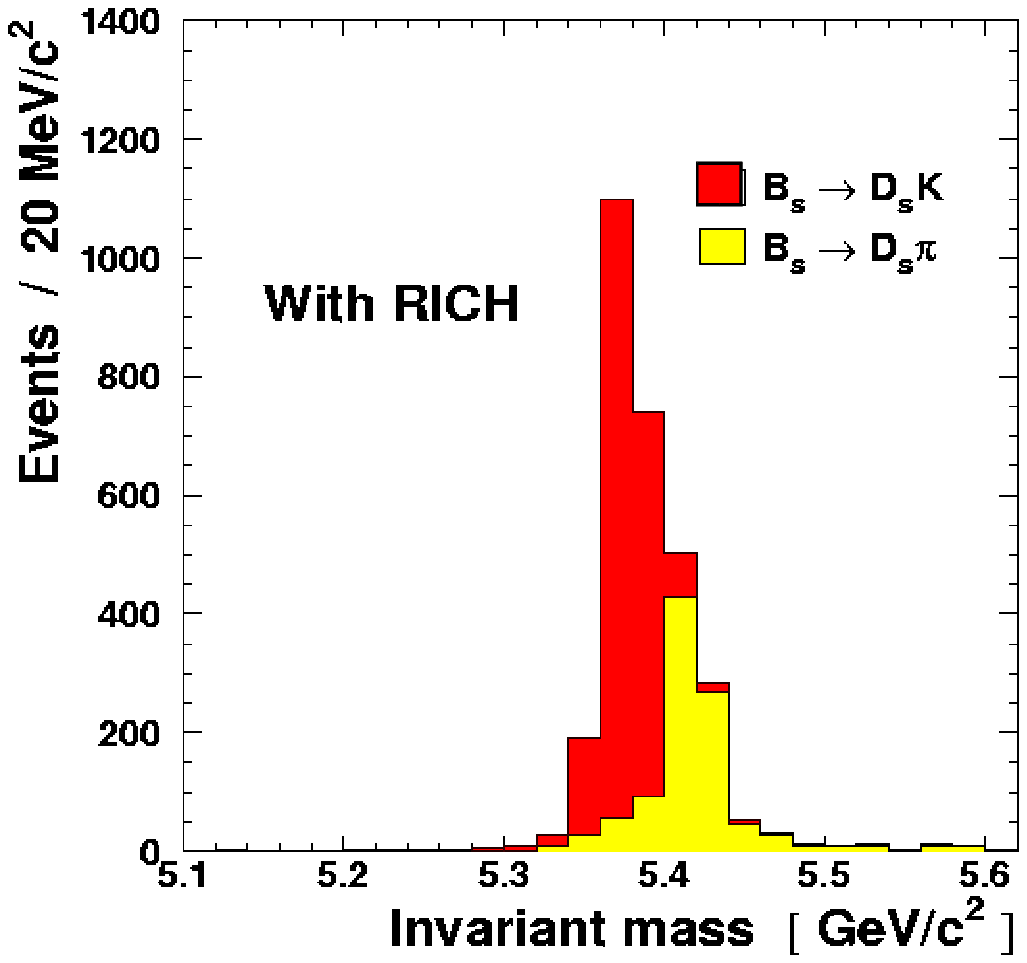}
  \caption{Without particle identification from the Cherenkov
    detectors in a simulation from \lhcb~(top) the $\Bs \to D_s^\mp
    K^\pm$ signal is drowned by the $\Bs \to D_s^- \pip$ decay. With
    particle identification~(bottom), the signal is dominant compared
    to the background.}
  \label{fig:DsK}
\end{figure}

To make hadronic final states useful for \CP violation studies it is
required that we can distinguish pions and kaons very well. A good
example is for the $\Bs \to D_s^\mp K^\pm$ decay to be used for the
extraction of the angle $\gamma$. The decay $\Bs \to D_s^- \pip$ is
expected to have a branching fraction 15 times larger than the kaon
decay thus drowning the $\Bs \to D_s^\mp K^\pm$ signal without any
particle identification. In figure~\ref{fig:DsK} we illustrate the
particle identification capability of \lhcb\ to isolate the $\Bs \to
D_s^\mp K^\pm$ signal. For the two-body \B meson decays the kaon-pion
separation is also essential for the extraction of the angle $\gamma$
from the individual measurements of $\Bd \to \pip \pim$ and $\Bs \to
\Kp \Km$ decays.

Both \lhcb\ and \btev\ use Ring Imaging Cherenkov~(RICH) detectors for
pion and kaon identification. For a given radiator the effective
momentum range is limited from below by the onset of Cherenkov
radiation for the pions and from above when the kaon Cherenkov angle
saturates causing the rings created by kaons and pions to have the
same radius. This means that more than one radiator is required.
\atlas\ and \cms\ have a very limited ability for kaon-pion
separation.

Kaon identification is one of the dominant sources for flavour
tagging. This can either be through identifying the charge of a kaon
from the decay of the other \B created in the event or for the tagging
of \Bs decays from the decay $\Bs^{**} \to \Bs \Kp$. In addition to
kaon-pion separation it is important to reduce the contamination of
the kaon tagging sample with protons. The \lhcb\ experiment use an
aerogel radiator with refractive index of 1.03 in addition to the two
gas radiators to provide kaon and proton separation down to around 2~\gevc.

\section{Physics reach}
\label{sec:Physics}
The aim of giving numbers for the physics reach before the start-up of
experiments is to assure that the detector design is able to give the
promised results in a selection of channels that are thought to be
representative of the physics that will be of interest in 2007 and
beyond. In the same spirit we will here only give a few examples of
the physics that can be addressed at the future \B factories. No
attempt has been made to be comprehensive. The numbers in this section
are from~\cite{Ball:2000ba,LHCb:TP,BTeV:Proposal,CMS:DAQTDR} and later
conference updates.

In general the strategy for the experiments will be to make several
measurements that in independent ways test the Standard Model.

As an example the $\beta$ measurement from $\Bd \to \jpsi\KS$ and the
measurement of $\Vtd/\Vcb$ through \Bd mixing are both sensitive to
new physics contributions in the \Bd loop diagram. On the other hand a
measurement of $\gamma$ from the $\Bs \to D_s^\mp K^\pm$ decays and of
$\Vub/\Vcb$ from the branching fractions of $\Bd \to h_u \ell \nu$ and
$\Bd \to h_c X$ decays only probes processes at the tree level and are
as such not expected to be sensitive to new physics.

\begin{figure}[htbp]
  \centering
  \includegraphics[width=\linewidth]{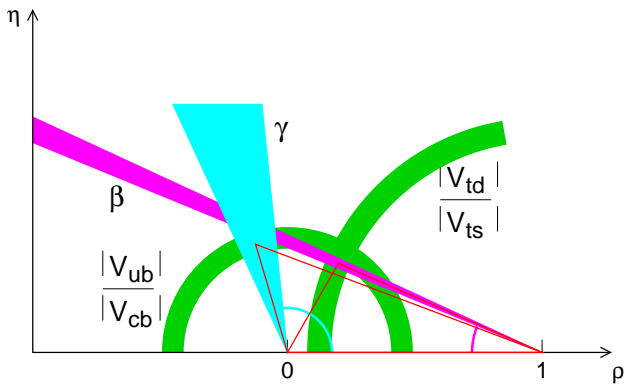}
  \caption{An illustration of how a mismatch is possible between
    measurements of the unitarity triangle which are sensitive to new
    physics and measurements which are not.}
  \label{fig:mismatch}
\end{figure}

This means we can form a \emph{standard triangle} from the angle
$\gamma$ and $\Vub/\Vcb$ measurements and a \emph{new physics
  triangle} from $\beta$ and $\Vtd/\Vcb$. If these two triangles do
not share the same apex we have a sign of new physics. The principle
is illustrated in figure~\ref{fig:mismatch}.

\subsection{\boldmath \Bs mixing and the \CP angle $\delta\gamma$}
\label{sec:Bsmix}
Within the Standard Model the weak phase $\phi_s$ in \Bs mixing is
given by the small value $-2 \delta\gamma \equiv - 2 \lambda^2 \eta$.
This means that new physics could easily show up as a larger value of
\CP violation in a decay like $\Bs \to \jpsi \phi$ which is equivalent
to the $\Bd \to \jpsi \KS$ decay for the measurement of the phase
$2\beta$ in \Bd mixing. The good calorimetry of \btev will also allow
a measurement of $\phi_s$ in the decay $\Bs \to \jpsi
\eta^{(\prime)}$.  The precision in the angle $\phi_s$ will depend on
how fast the oscillation frequency is for \Bs mixing but will in
general be around 0.02.

\subsection{\boldmath Extraction of the angle $\gamma$}
\label{sec:gamma}
The extraction of the \CP angle $\gamma$ is one of the main purposes
for the hadronic \B factories. The decay $\Bs \to D_s^\mp K^\pm$ is
sensitive to the angle $\phi_s + \gamma$, where the $\phi_s$ part
comes from \Bs mixing and the $\gamma$ part from the phase of $V_{ub}$
in the tree level decay.  If new physics contributes to $\phi_s$ it
will be the same contribution as for the direct measurement of
$\phi_s$ and as such will not interfere with a clean measurement of
$\gamma$ from the tree level decay. The decay $\Bd \to \Dstarpm \pimp$
is the equivalent decay for \Bd but suffers from the problem that one
of the interfering decays is doubly-Cabibbo-suppressed with respect to
the other; the increased statistics in this channel due to the large
branching fraction will more or less cancel the deterioration in
sensitivity from the doubly-Cabibbo-suppressed amplitude leading to a
similar overall sensitivity to $\gamma$.

The angle $\gamma$ can also be extracted with high precision from a
comparison of $\Bd \to \pip pim$ and $\Bs \to \Kp
\Km$~\cite{Fleischer:1999pa}. This method is sensitive to new phases
introduced in the penguin decays and as such might not measure the
Standard Model value of $\gamma$. In table~\ref{tab:gamma} we
summarise the statistical samples available with one year of data for
a $\gamma$ measurement.
\begin{table}[htbp]
  \centering
  \begin{tabular*}{\linewidth}[c]{@{\extracolsep{\fill}}lccc}
     Decay channel              & \lhcb & \btev & Possible new phases? \\
     \hline
     $\Bs \to D_s^\mp K^\pm$    & 8~k   & 7.5~k & No  \\
     $\Bd \to \Dstarpm \pimp$   & 650~k &       & No  \\
     $\Bd \to \pip \pim$        & 27~k & 15~k  & \multirow{2}{0.5cm}{Yes} \\
     $\Bs \to \Kp \Km$          & 38~k & 19~k  \\
  \end{tabular*}
  \caption{The expected statistics in one year at design luiminosity
     for different channels for the extraction of the angle $\gamma$.
     All have the potential for an accuracy of 10\degrees or better
     in $\gamma$.}
  \label{tab:gamma}
\end{table}

\subsection{Rare decays}
\label{sec:rare}
The predicted branching fraction from the Standard Model for the $\Bd
\to \mumu$ decay is around $10^{-10}$ and for the $\Bs \to \mumu$
decay around $4\times 10^{-9}$. As the decays have to be mediated
through a loop there is the possibility that particles from new
physics will participate and dramatically increase the decay rate.
Table~\ref{tab:rare} summarises the expected number of reconstructed
events given the Standard Model decay rate. As it can be seen a
measurement of the $\Bd \to \mumu$ decay will be marginal even at high
luminosity for \cms\ and \atlas\ as there may well be significant
background. The background levels are almost impossible to evaluate
from simulations due to the very large rejection factors required.
\begin{table}[htbp]
  \centering
  \begin{tabular*}{\linewidth}[c]{@{\extracolsep{\fill}}lcccc}
     Decay channel          & \atlas & \cms & \btev & \lhcb     \\
     \hline
     $\Bd \to \mumu$        & 14     & 4.1  & 3     &           \\
     $\Bs \to \mumu$        & 27     & 21   & 18    & 30        \\
     $\B\ \to \Kstarz\mumu$ & 2~k    & 12~k & 8~k   & 13~k      \\
  \end{tabular*}
  \caption{The expected reconstructed rate for Standard Model
     production of rare decays. Statistics are after 3 years at the
     nominal luminosity for \lhcb\ and \btev, and 3 years of low
     luminosity ($\lum = 10^{33}~\cmsqs$) for \atlas\ and \cms\
     (except $\Bd \to \mumu$ which is one year at high luminosity
     ($\lum = 10^{34}~\cmsqs$) for \atlas\ and \cms).}
  \label{tab:rare}
\end{table}

\section{Systematics}
\label{sec:systematic}
In the experiments we need to control everything that can give a
flavour asymmetry and fake \CP violation. There are several
penitential sources for a flavour asymmetry:
\begin{itemize}
\item Since LHC is a proton-proton machine the angular distributions
  and relative ratios of the different types of \B and \Bb hadrons
  will be different at the percent level which is larger than some of
  the effects we want to measure.
\item The tracking efficiency for positive and negative particles will be
  different due to the magnetic dipole field (positive and negative
  particles go through different parts of the detector).
\item Particle identification will be different for \Kp and \Km due
  the the difference in nuclear interaction length.
\item The flavour tagging will be different due to asymmetries in both
  the efficiency and mistag rates.
\end{itemize}
All these effects should be measured and corrected from analysing the
data. Separate control channels should be found for each of the
different types of hadrons and care should be taken that there is no
expected direct \CP violation in the control channels.

\section{Conclusions}
\label{sec:conclusion}
The \B factories at hadronic colliders will in the future provide
statistics of the order of $10^{12}$ \bbbar pairs per year. A
sophisticated trigger is required to reduce the background from the
much larger production of minimum bias events and to select the
specific \B decays of interest.

The \lhcb\ and \btev\ detectors are optimised to cover a wide range of
(semi)-leptonic and hadronic decays with high efficiency. The \atlas\ 
and \cms\ experiments will be competitive in channels with a muon pair
in the final state which could be of great interest in the detection
of rare \B decays.

The experiments should together be able to make comprehensive
measurements of the \CP violating effects in the quark sector.
Hopefully we will from this see that the single \CP violating phase of
the Standard Model is no longer sufficient to explain all the data and
that new phases from New Physics are required.

\section{Acknowledgements}
\label{sec:ack}
I would like to thank the organisers of the workshop for providing a
good atmosphere for discussions. Also I would like to thank
P.~Eerola~(\atlas), F.~Palla, A.~Starodumov~(\cms) and
L.~Moroni~(\btev) for providing me with information and corrections on
the other experiments. Naturally any remaining errors are my
responsibility.

\bibliographystyle{h-physrev3}
\bibliography{egede}

\end{document}